\documentclass[prb,aps,twocolumn,showpacs,superscriptaddress, amsmath,amssymb,10pt,letterpaper]{revtex4-1}
\usepackage{comment}
\usepackage{xcolor}
\usepackage{soul}

\usepackage{graphicx}
\usepackage{dcolumn}
\usepackage{bm}
\usepackage{dsfont}
\usepackage{textcomp} 

\newcommand{\nn}{\nonumber}

\def\bea{\begin{equation}\begin{aligned}}
\def\eea{\end{aligned}\end{equation}}

\begin{document}
\large
\title{Spin Diffusion in Spin Glasses Requires Two Magnetic Variables, $\vec{M}$ and $\vec{m}$}
\author{Chen Sun}
\affiliation{ School of Physics and Electronics, Hunan University, Changsha 410082, China}
\author{Wayne M. Saslow}
\email{wsaslow@tamu.edu}
\affiliation{ Texas A\&M University, College Station, Texas, 77843, U.S.A.}
\begin{abstract}
Experiment has established that spin-glasses can support a steady-state spin current $\vec{j}_{i}$.  However, the accepted theory of spin glass dynamics permits oscillations but no steady-state spin current.  Onsager's irreversible thermodynamics implies that the spin current is proportional to the gradient of a magnetization.  We argue, however, that the magnon distribution function associated with the local equilibrium magnetization $\vec{M}$ cannot diffuse because it represents $10^{23}$ variables.  We therefore invoke the non-equilibrium magnetization $\vec{m}$, which in spintronics is called the {\it spin accumulation}.  Applying the theory of irreversible thermodynamics we indeed find that it predicts spin diffusion, and we consider other experimental consequences of the theory, including a wavelength-dependent coupling between the reactive and the diffusive degrees of freedom.

\date{\today}

\end{abstract}

\maketitle


\section{Introduction}

Spintronics, i.e. control and manipulation of spin degrees of freedom,\cite{ZFD-RMP,Wolf05} especially spin currents, has received great attention in recent years.   Various magnetic materials including paramagnets, ferromagnets and anti-ferromagnets has been used to realize spintronics. The present work considers spintronics in spin glasses, which dates to 2011.  At that time Iguchi {\it et al} studied spin current absorption in the spin glass AgMn using the spin pumping method, and proposed spin current injection as a method to investigate spin glass properties\cite{Iguchi10}.

More recently, Wessenberg {\it et al} presented evidence for steady-state spin currents traversing an amorphous sample of the magnetic insulator yttrium iron garnet (YIG).\cite{NatPhysWuZink17}  In their work x-ray diffraction on a 200 nm sputtered sample indicated no medium-range or long-range order.  From this it was inferred that the system was polycrystalline and perhaps, for lengths scales larger than 100 nm, effectively a spin glass.\cite{Mydosh15}  For these thin films they found spin currents much larger than for crystalline YIG. 
This experiment has motivated us to develop a theory for spin currents in spin glasses.  

Unfortunately, further experiment is needed in this area, such as -- all other variables fixed -- a  study of the dependence of the spin current on sample thickness.  For dc spin current such a study has been performed on an antiferromagnet, which like a spin glass has no net magnetization.  That study found two characteristic decay lengths.\cite{WangHammel14}  (Ref.~\onlinecite{Rezende16} uses a kinetic theory to discuss spin diffusion for insulating antiferromagnets.) For a spin glass we expect that, due to its isotropic nature, there would be only one characteristic decay length.

At the microscopic level we expect that in a spin glass the equilibrium spin orientations vary rapidly from site to site, due to the frustrating effect of competing exchange interactions.  This is illustrated explicitly in the simulations of Walker and Walstedt.\cite{WalkerWalstedt77,WalkerWalstedt80}  Previous theories of the macroscopic dynamics of spin glasses included as macroscopic variables the magnetization $\vec{M}$ and an orientation $\vec{\theta}$.\cite{HalperinSaslow77,Andreev78}  The variable $\vec{\theta}$ represents a local rotation of the noncollinear magnetic order within the spin glass.  
For a spin-glass sample with no magnetization in zero external field $H$ (i.e no remanence), the local spins tip to give a local net magnetization $\vec{M}$.  However, these theories give no spin current $\vec{j}_{i}$, where $i$ is a spacial index.  Note that a spin-glass repeatedly prepared in the same field $\vec{H}$ and the same sample orientation (thus the same local anisotropies) will in each case go into distinct microscopic states.  However, all of these microscopic states will have the same macroscopic properties.  In equilibrium we take $\vec{\theta}=\vec{0}$.

The present work applies the spintronics idea\cite{ValetFert93,ZhangLevyFert02} that spin currents are associated with the non-equilibrium magnetization $\vec{m}$ known as the {\it spin accumulation}.
To our knowledge there is no previous theoretical work on spin currents (i.e, spin diffusion) in spin glasses, and the present work is directed at filling this gap to provide a framework to understand the experimental results in Ref.~\onlinecite{NatPhysWuZink17}.  

As background, Sect.~II discusses the spin accumulation $\vec{m}$, Sect.~III discusses the energy density when $\vec{m}$ is included, and Sect.~IV discusses the corresponding thermodynamics.   Sect.~V uses irreversible thermodynamics to obtain the equations of motion.  As a check, Sect.~VI obtains the pure spin waves, where $\vec{m}$ is neglected, and Sect.~VII obtains the pure spin diffusion modes, where $\vec{M}$ and $\vec{\theta}$ are neglected.  Sect.~VIII discusses the fully coupled modes, and Sect.~IX discusses the weakly coupled modes, where either a spin wave is accompanied by a small amplitude for spin diffusion, or where spin diffusion is accompanied by a small amplitude spin wave.  Sect.~X presents a summary and discussion that relates the present work, for insulating spin glasses, to spin glasses with a conductor like Cu that has been doped with Mn.  An Appendix discusses the RKKY interaction in both nuclear and electron spin systems.

\section{Spin Accumulation}
For magnetic insulators with magnon modes labeled by $\alpha$, in local equilibrium we write the thermal occupation number $n^{l.e.}_{\alpha}$.  A deviation from equilibrium $\delta n_{\alpha}$ of the thermal occupation number $n_{\alpha}$ can give a local value for the non-equilibrium magnetization, which we identify with $\vec{m}$ rather than $\vec{M}$.  $\vec{M}$ may correspond to some equilibrium state of the system, although perhaps not for the actual $\vec{H}$, as can happen if $\vec{H}$ is rapidly changed.  
This approach permits a near-equilibrium spin glass to support spin diffusion in terms of a diffusive non-reactive magnetic variable, $\vec{m}$, while permitting $\vec{M}$ to remain a non-diffusive reactive variable.

Translated back into the language of site localized spins, we distinguish between the local value of the equilibrium part of the spin $\vec{S}_{i}$ and the non-equilibrium part of the spin $\vec{s}_{i}$.  Summing the former over spins gives the local equilibrium $\vec{M}$ and summing over the latter gives the local non-equilibrium $\vec{m}$.  Even before the term {\it spintronics} was coined, the idea of a local non-equilibrium $\vec{m}$ was employed by Dyakonov and Perel to predict, for semiconductors with a spin-orbit interaction, what later became known as the spin Hall effect and the inverse spin Hall effect.\cite{DPJETPLett71,DPPhysLett71}.  They used the term {\it accumulation of spin}.

We recently applied the idea of spin accumulation $\vec{m}$ to the longitudinal magnetization $M$ of a ferromagnet.  There we argued that spin diffusion of $M$ from position A to position B takes place indirectly, first at A via conversion of $M$ to $m$, then by diffusion of $m$ from A to B, and finally by conversion at B of $m$ to $M$. \cite{SasSunXu22}
We also argued, implicitly, that any non-conserved order parameter (call it $Y$; it could be a superconducting order parameter), being encoded in a statistical distribution function of $10^{23}$ variables, is unable to diffuse, and therefore diffusion of $Y$ from A to B occurs indirectly, first at A by conversion from $Y$ to the ``$Y$-accumulation'' $y$, then by diffusion of $y$ from A to B, and finally by conversion at B from $y$ to $Y$.

It is essential to recognize that spin glasses, with no magnetization and isotropic macroscopic properties, have a different symmetry than do ferromagnets. Ferromagnets have a single longitudinal magnetic variable and two transverse magnetic variables.  On the other hand, spin glasses have three rotationally equivalent magnetic variables.  In the absence of spin diffusion and anisotropy, ferromagnets have a decaying, non-propagating longitudinal mode, and two degenerate transverse modes with quadratic dependence on the wavevector $k$.  On the other hand, under the same conditions spin glasses have three degenerate modes with a linear dependence on $k$.  This is much more like that for antiferromagnets, which, however, have a more complex symmetry than do spin glasses.  In part for that reason the present work does not consider spin diffusion in antiferromagnets.

The simplest system supporting a spin current is a paramagnet, with spin current proportional to the gradient of the non-equilibrium magnetization $\vec{m}$.\cite{Aranov76,JohnsonSilsbee87a,Wyder87,ValetFert93}  Both a spin diffusion constant $D$ and a spin relaxation time $\tau$ are present, and give a characteristic and dissipative spin decay length $l_{s}\sim(D\tau)^{1/2}$.  In the end, for a spin glass with $\vec{M}$, $\vec{\theta}$, and in addition $\vec{m}$, we find a similar decay length, but complicated by dimensionless factors related to ratios of additional decay times and ratios of the susceptibilities of $\vec{M}$ and $\vec{m}$.

\section{Energy Density}
We consider that a spin glass is described by a set of local atomic magnet orientations that, when only exchange interactions are included, can be uniformly rotated by an infinitesimal angle $\vec{\theta}$ to produce an inequivalent microscopic state with no change in energy.\cite{WalkerWalstedt77,WalkerWalstedt80,HalperinSaslow77}  For a given microscopic configuration, when weak anisotropy interactions are included the spins will locally re-orient slightly to give a local minimum that we will characterize by $\vec{\theta}=\vec{0}$.

If cooled in zero external field (zfc), the system will develop no magnetization, and when a non-zero $\vec{H}$ is applied, in equilibrium it will develop a {\it magnetization} $\vec{M}$ with
\begin{equation}
\vec{M}=\chi_{M} \vec{H}\quad \rm (zfc),
\label{MH=0}
\end{equation}
where $\chi_{M}$ is the susceptibility of $\vec{M}$.

If a spin-glass is prepared in a finite external field $\vec{H}_{fc}$ (field-cooling), then on removing $\vec{H}_{fc}$ there is a remanent magnetization $M_{0}$ along $\vec{H}_{c}$.  When an additional field $\vec{H}$ is applied along $\vec{H}_{fc}$ we have $\vec{M}$ along $\vec{H}_{fc}$.  Therefore $\vec{M}_{0}$ and $\vec{H}$ are collinear, so we may write the scalar equation
\begin{equation}
M=M_{0}+\chi_{M} H.\quad
\label{MHne0}
\end{equation}
A three-dimensional anisotropy (see below), due to single ion anisotropy or Dzyaloshinsky-Moriya anisotropy,\cite{LevyMPFert82,HSH82,WMS83} tends to align $\vec{M}_{0}$ with the direction of $\vec{H}_{fc}$, and defines a 3d baseline set of coordinates relative to which the spin can be rotated.  For small rotations we employ the 3d angle $\vec{\theta}$.   Compared to the saturation magnetization, $M_{0}$ typically is small, so that the system is only slightly affected by it.

\subsection{Energy Density with $\vec{M}$ and $\vec{\theta}$}
In the absence of $\vec{m}$, following previous work we employ variables $\vec{M}$ and $\vec{\theta}$.  We assume an energy density of the form
\begin{equation}
\frac{\varepsilon_{0}}{\mu_{0}}=\frac{(\vec{M}-\vec{M}_{0})^{2}}{2\chi_{M}}-\vec{M}\cdot\vec{H}+\frac{K}{2}\vec{\theta}^{\,2}+\frac{\rho_{s}}{2}(\partial_{i}\vec{\theta})^{2},
\label{energy0}
\end{equation}
where $\mu_{0}$ is the vacuum permeability. The first term permits a frozen-in remanent magnetization (common in spin-glasses) and the second term is the Zeeman energy. The third term, with anisotropy constant $K$, represents the macroscopic anisotropy energy, which is minimized for $\vec{\theta}=\vec{0}$.  
The last term, with spin stiffness coefficient $\rho_{s}$, represents the increase in microscopic exchange energy for a {\it nonuniform} rotation $\vec{\theta}$.

In the absence of $K$, to produce a twist in $\vec{\theta}$ alone would require the spin analog of what, in numerical simulations for atoms, the distinguished computational physicist Aneesur Rahman called ``ether pegs.''\cite{Rahman85}  With ``spin ether pegs'' to twist the spin texture, the new twisted equilibrium state will have characteristic twist length $l_{n}=(\rho_{s}/K)^{1/2}$.   This non-dissipative length does not relate to the dissipative spin decay length, associated with both spin decay and spin diffusion, that we will find below.

\subsection{Energy Density with $\vec{m}$}
When spin accumulation is included, we append the additional energy density
\begin{eqnarray}
\frac{\varepsilon_{1}}{\mu_{0}}&=&\frac{m^{2}}{2\chi_{m}}-\vec{m}\cdot\vec{H}-\lambda_{M}\vec{m}\cdot(\vec{M}-\vec{M}_{0}),
\label{energy1}
\end{eqnarray}
where $\chi_{m}$ is the susceptibility of $\vec{m}$. Here the first term has its origin in exchange, and the second term is a Zeeman interaction.  The last term, involving the effective exchange constant $\lambda_{M}$, is of a statistical nature.  Enforcing the equilibrium conditions $\vec{M}_{eq}=\vec{M}_{0}+\chi_{M}\vec{H}$ and $\vec{m}_{eq}=\vec{0}$ will constrain the exchange constant $\lambda_{M}$.

\section{Thermodynamics}
We now give the thermodynamic differential energy density $\varepsilon=\varepsilon_0+\varepsilon_1$ more generally, in terms of the appropriate thermodynamic variables.  We have, with entropy density $s$ and temperature $T$,
\begin{equation}
d\varepsilon=Tds-\mu_{0}\vec{H}^{*}\cdot d\vec{M}-\mu_{0}\vec{h}^{*}\cdot d\vec{m}-\vec{\Gamma}\cdot d\vec{\theta}-\vec{\Gamma}_{i}\cdot d(\partial_{i}\vec{\theta}),
\label{deps0}
\end{equation}
where $\vec{H}^{*}$, $\vec{h}^{*}$ and $\vec{\Gamma}_{i}$ are thermodynamic conjugates (effective fields) associated with $\vec{M}$, $\vec{m}$, and $\partial_{i}\vec{\theta}$.  When \eqref{energy0} applies, $\vec{\Gamma}=-K\vec{\theta}$, and $\vec{\Gamma}_{i}=-\rho_{s}\partial_{i}\vec{\theta}$.

We will require that the conditions $\vec{H}^{*}_{eq}=\vec{0}$ and $\vec{h}^{*}_{eq}=\vec{0}$ recover the equilibrium conditions $\vec{M}=\vec{M}_{0}+\chi_{M}\vec{H}$ and $\vec{m}=\vec{0}$.

The variational derivative of the energy with respect to $\vec{\theta}$ defines the net torque density
\begin{equation}
\vec{\Gamma}'\equiv \vec{\Gamma}-\partial_{i}\vec{\Gamma}_{i},
\label{Gamma'}
\end{equation}
and from \eqref{energy0}, we have
\begin{equation}
\vec{\Gamma}'=-K\vec{\theta}+\rho_{s}\nabla^{2}\vec{\theta}.
\label{Gamma'2}
\end{equation}
This permits us to simplify \eqref{deps0} to, on neglecting a pure divergence term,
\begin{equation}
d\varepsilon=Tds-\mu_{0}\vec{H}^{*}\cdot d\vec{M}-\mu_{0}\vec{h}^{*}\cdot d\vec{m}-\vec{\Gamma}'\cdot d\vec{\theta}.
\label{deps1}
\end{equation}

\subsection{The Effective Field $\vec{H}^{*}$}
We define the effective field $\vec{H}^{*}$ by 
\begin{equation}
\vec{H}^{*}\equiv-\frac{\partial}{\partial\vec{M}}\frac{\varepsilon}{\mu_{0}}=\vec{H}-\frac{\vec{M}-\vec{M}_{0}}{\chi_{M}}+\lambda_{M}\vec{m}.
\label{H*1}
\end{equation}
$\vec{H}^{*}_{eq}=\vec{0}$ is consistent with the equilibrium conditions.

We now introduce the magnetization deviation from local equilibrium, $\delta \vec{M}$, via
\begin{equation}
\delta \vec{M}\equiv-\chi_{M}\vec{H}^{*}=\vec{M}-\vec{M}_{0}-\chi_{M}(\vec{H}+\lambda_{M}\vec{m}).
\label{delM1}
\end{equation}

\subsection{The Effective Field $\vec{h}^{*}$}
We define the effective field $\vec{h}^{*}$ by 
\begin{equation}
\vec{h}^{*}\equiv-\frac{\partial}{\partial\vec{m}}\frac{\varepsilon}{\mu_{0}}=\vec{H}-\frac{\vec{m}}{\chi_{m}}+\lambda_{M}(\vec{M}-\vec{M}_{0}).
\label{h*1}
\end{equation}
We can make this consistent with the equilibrium conditions if we take
\begin{equation}
\lambda_{M}=-\frac{1}{\chi_{M}},
\label{lambdaM}
\end{equation}
so
\begin{equation}
\vec{h}^{*}\equiv-\frac{\partial}{\partial\vec{m}}\frac{\varepsilon}{\mu_{0}}=\vec{H}-\frac{\vec{m}}{\chi_{m}}-\frac{(\vec{M}-\vec{M}_{0})}{\chi_{M}}.
\label{h*2}
\end{equation}

We now introduce the spin accumulation deviation from local equilibrium, $\delta \vec{m}$, via
\begin{equation}
\delta \vec{m}\equiv-\chi_{m}\vec{h}^{*}=\vec{m}-\frac{\chi_{m}}{\chi_{M}}(\vec{M}-\vec{M}_{0}-\chi_{M}\vec{H}).
\label{delm0}
\end{equation}

\subsection{Deviations from Local Equilibrium}
For completeness we rewrite $\vec{H}^{*}$ and $\delta \vec{M}$ as
\begin{equation}
\vec{H}^{*}=\vec{H}-\frac{\vec{M}-\vec{M}_{0}}{\chi_{M}}-\frac{\vec{m}}{\chi_{M}},
\label{H*2}
\end{equation}
\begin{equation}
\delta \vec{M}=\vec{M}-\vec{M}_{0}-\chi_{M}\vec{H}+\vec{m}.
\label{delM2}
\end{equation}

With \eqref{lambdaM} and the definition
\begin{equation}
\xi\equiv\frac{\chi_{m}}{\chi_{M}},
\label{xi}
\end{equation}
we may write, in defining $\Delta\vec{m}$ and $\Delta\vec{M}$,
\begin{eqnarray}
\delta \vec{m}&\equiv& -\chi_{m}\vec{h}^{*}\cr
&=&\vec{m}+\frac{\chi_{m}}{\chi_{M}}(\vec{M}-\vec{M}_{0}-\chi_{M}\vec{H})\cr
&\equiv&\Delta \vec{m}+\xi \Delta \vec{M},
\label{delm_L}\\
\delta \vec{M}&\equiv& -\chi_{M}\vec{H}^{*}\cr
&=&(\vec{M}-\vec{M}_{0}-\chi_{M}\vec{H})+\vec{m}\cr
&\equiv&\Delta \vec{M}+\Delta \vec{m}.
\label{delM_L}
\end{eqnarray}

For uniform systems the equilibrium conditions $\vec{H}^{*}=\vec{0}$ and $\vec{h}^{*}=\vec{0}$ yield equations like those for two interpenetrating and interacting paramagnets, so for $\vec{H}\ne\vec{0}$ both $\vec{M}\ne\vec{0}$ and $\vec{m}\ne\vec{0}$.

\section{Irreversible Thermodynamics and Equations of Motion}
We now employ Onsager's irreversible thermodynamics to study the dynamics of this system, including dissipation.  For that we first write down the conservation laws and equations of motion for the thermodynamic variables.  We then require that, if \eqref{deps0} holds initially, then it holds in the future, subject to the non-decreasing nature of $s$.  
We take $\vec{M}$ to be non-diffusive because it represents an equilibrium distribution.  However, $\vec{m}$ can have a diffusive flux term.

This system has various unknown dissipative fluxes $j$ and dissipative sources $R$.  We assume that: \\
(a) energy has only a flux $j^{\varepsilon}_{i}$, \\
(b) entropy has a flux $j^{s}_{i}$ and a source $R_{s}\ge0$, \\
(c) $\vec{M}$ has only a source $\vec{R}_{M}$ (it has no flux because we believe that a macroscopic variable should not be able to diffuse), and is driven by torque both from the effective field and from the anisotropy, or $-\gamma\vec{M}\times\vec{H}^{*}-\gamma\vec{\Gamma}'$, \\
(d) $\vec{m}$ has a flux $\vec{j}^{m}_{i}$ and a source $\vec{R}_{m}$, and is driven by $-\gamma\vec{m}\times\vec{h}^{*}$, with no lattice torque $\vec{\Gamma}'$ analogous to $\vec{\Gamma}$,\\
(e) $\vec{\theta}$ has no flux but a source $\vec{R}_{\theta}$ and is driven by an unknown $\vec{\omega}$.  We expect that $\vec{\omega}=\gamma\mu_{0}\vec{H}^{*}$, because the atomic moments in a spin glass precess in the effective field.

Thus, with gyromagnetic ratio due to electrons (properly as $-\gamma<0$) we take
\begin{eqnarray}
\partial_{t}\varepsilon+\partial_{i}j^{\varepsilon}_{i}&=&0, \label{dedt0}\\
\partial_{t}s+\partial_{i}j^{s}_{i}&=&R_{s}\ge0, \label{dsdt0}\\
\partial_{t}\vec{M}&=&-\gamma\mu_{0}\vec{M}\times\vec{H}^{*}-\gamma\vec{\Gamma}'+\vec{R}_{M}, \quad \label{dMdt0}\\
\partial_{t}\vec{m}+\partial_{i}\vec{j}^{m}_{i}&=&-\gamma\mu_{0}\vec{m}\times\vec{h}^{*}+\vec{R}_{m}, \label{dmdt0}\\
\partial_{t}\vec{\theta}&=&\vec{\omega}+\vec{R}_{\theta}. \label{dthetadt0}
\end{eqnarray}

Using \eqref{deps1} and the above equations gives
\begin{eqnarray}
-\partial_{i}j^{\varepsilon}_{i}&=&-T\partial_{i}j^{s}_{i}+TR_{s}{-}\mu_{0}\vec{H}^{*}\cdot\vec{R}_{M}{-}\mu_{0}\vec{h}^{*}\cdot\vec{R}_{m}\cr
&&+\mu_{0}\vec{h}^{*}\cdot\partial_{i}\vec{j}^{m}_{i}{-}\vec{\Gamma}'\cdot(\vec{\omega}{-}\gamma\mu_{0}\vec{H}^{*}+\vec{R}_{\theta}).\qquad
\label{deps2}
\end{eqnarray}

Then
\begin{eqnarray}
0\le TR_{s}&=&-\partial_{i}[j^{\varepsilon}_{i}-Tj^{s}_{i}
+\mu_{0}\vec{h}^{*}\cdot\vec{j}^{m}_{i}]\cr
&&-j^{s}_{i}\partial_{i}T+\mu_{0}\vec{H}^{*}\cdot\vec{R}_{M}\cr
&&+\mu_{0}\vec{h}^{*}\cdot\vec{R}_{m}+{\mu_0}\vec{j}^{m}_{i}\cdot\partial_{i}\vec{h}^{*}\cr
&&{+}\vec{\Gamma}'\cdot(\vec{\omega}{-}\gamma\mu_{0}\vec{H}^{*}+\vec{R}_{\theta}).
\label{Rs1}
\end{eqnarray}

For $R_{s}$ to be non-negative, we take \\
(a) the entropy flux to be
\begin{equation}
j^{s}_{i}=-\frac{\kappa}{T}\partial_{i}T,
\label{j^s}
\end{equation}
where $\kappa\ge0$ is the thermal conductivity, with units of entropy density 
diffusion constant; \\
(b) the $\vec{M}$ and $\vec{m}$ source terms to satisfy
\begin{eqnarray}
\vec{R}_{M}=-\frac{\delta \vec{M}}{\tau_{M}}+\frac{\delta \vec{m}}{{\tau_{mM}}}
=\frac{\chi_{M}}{\tau_{M}}\vec{H}^{*}-\frac{\chi_{m}}{\tau_{mM}}\vec{h}^{*},
\label{RM}\\
\vec{R}_{m}=-\frac{\delta \vec{m}}{\tau_{m}}+\frac{\delta \vec{M}}{{\tau_{Mm}}}
=\frac{\chi_{m}}{\tau_{m}}\vec{h}^{*}-\frac{\chi_{M}}{\tau_{Mm}}\vec{H}^{*},
\label{Rm}
\end{eqnarray}
with self-explanatory positive relaxation times; \\
(c) the magnetization flux to come only from diffusion of $\vec{m}$, as in
\begin{equation}
\vec{j}^{m}_{i}= \frac{D_{m}}{\chi_{m}}\partial_{i}\vec{h}^{*}=-D_{m}\partial_{i}\delta\vec{m},
\label{j^m}
\end{equation}
where $D_{m}\ge0$ is a diffusion coefficient, with units of velocity times distance; \\
(d) the reactive part of the $\dot{\vec{\theta}}$ driving term to be
\begin{equation}
\vec{\omega}= \gamma\mu_{0}\vec{H}^{*};
\label{omega}
\end{equation}
(e) the dissipative part of the $\dot{\vec{\theta}}$ driving term to be dissipative
\begin{equation}
\vec{R}_{\theta}= \alpha\gamma\vec{\Gamma}',
\label{Rtheta}
\end{equation}
where $\alpha\ge0$ is a dissipation coefficient, with units of inverse magnetization.

We also note the Onsager relation between cross-decay rates, which can be obtained by, in $TR_{s}$, equating the cross-terms in \eqref{RM} and \eqref{Rm}:
\begin{equation}
\frac{\chi_{m}}{\tau_{mM}}=\frac{\chi_{M}}{\tau_{Mm}}.
\label{Onsager}
\end{equation}

\section{Pure Spin Waves: Neglect $\vec{m}$}

In the absence of $K$, $\vec{H}$, dissipation, and $\vec{m}$, by \eqref{Gamma'2}  we have $\vec{\Gamma}'= \rho_{s}\nabla^{2}\vec{\theta}$.  Further, by \eqref{delM_L} we have $\vec{H}^{*}=-\Delta\vec{M}/\chi_{M}$, so
\begin{eqnarray}
\partial_{t}\vec{M}&=&-\gamma \rho_{s}\nabla^{2}\vec{\theta} ,
\label{dMdtK=0}\\
\partial_{t}\vec{\theta}&=& \gamma\mu_{0}(-\frac{\Delta\vec{M}}{\chi_{M}}).
\label{dthetadtK=0}
\end{eqnarray}
At fixed $\vec{M}_{0}$ and fixed $\vec{H}$, $\Delta\vec{M}=\vec{M}$, so combining these gives
\begin{equation}
\partial_{t}^{2}\vec{M}=\frac{\gamma^{2}\mu_{0}}{\chi_{M}}{\rho_s}\nabla^{2}\vec{M}.
\label{SGSW1}
\end{equation}
With $\vec{M}\sim e^{i(kx-\omega t)}$ we then have
\begin{equation}
\omega=vk, \quad v\equiv (\frac{\gamma^{2}\mu_{0}\rho_{s}}{\chi_{M}})^{1/2},
\label{SGSW2}
\end{equation}
the spin-wave mode given in Ref.~\onlinecite{HalperinSaslow77}.
This might appear in an experiment by imposing a field gradient with wavevector $k$ that oscillates at variable $\omega$ until there is a resonance.

If $K\ne 0$, then we may replace $K$ by $K'=K+\rho_{ss}k^{2}$.  In that case we have
\begin{equation}
\omega=(\omega_{0}^{2}+v^{2}k^{2})^{1/2}, \quad \omega_{0}\equiv \gamma(\frac{\mu_{0}K}{\chi_{M}})^{1/2}.
\label{SGSW3}
\end{equation}
This applies if $\omega_{0}$ is much larger than the relaxation time for $K$.\cite{Pastora83,WMS82}  If $\omega_{0}$ is much smaller than the relaxation time for $K$, then \eqref{SGSW2} applies.

\section{Pure Spin Diffusion: Neglect $\vec{M}$ and $\vec{\theta}$}
If a spin current $\vec{j}_{i}$ enters the system, and $\vec{M}$ and $\vec{\theta}$ can be neglected, then \eqref{dmdt0}, \eqref{Rm}, and \eqref{j^m} (on dropping the nonlinear term $\vec{m}\times\vec{h}^{*}$, and the term $\delta\vec{M}$), reduce to
\begin{equation}
\partial_{t}\vec{m}-{D_{m}}
\nabla^{2}\vec{m}=-\frac{1}{\tau_{m}}\vec{m}.
\label{j^mflow0}
\end{equation}
For the dc case we expect that $\vec{m}\sim e^{-qx}$, where the solution of the above equation gives $q^{-1}=l_{SG}={(D_{m}\tau_{m})^{1/2}}
$.  It would be of interest to test this by studying spin flow through samples of different thicknesses.

\section{Fully Coupled Modes}
If a field gradient is added, then $\vec{M}$ should develop a gradient, which might twist $\vec{\theta}$.  Let us consider how this might appear in an experiment.

For $\vec{H}=\vec{0}=\vec{M}_{0}$, by \eqref{delm_L} and \eqref{delM_L} we have
\begin{equation}
\delta\vec{M}=\vec{M}+\vec{m}, \quad \delta\vec{m}=\xi\vec{M}+\vec{m}.
\label{deltaMm}
\end{equation}

Then, using \eqref{dMdt0}, \eqref{RM}, and \eqref{H*2}, and neglecting precession,
\begin{equation}
\partial_{t}\vec{M}=-\gamma\vec{\Gamma}'-\frac{\delta \vec{M}}{\tau_{M}}+\frac{\delta \vec{m}}{{\tau_{mM}}};
\label{dMdtK=1}
\end{equation}
using \eqref{dthetadt0}, \eqref{omega}, \eqref{Rtheta}, and \eqref{delM2},
\begin{equation}
\partial_{t}\vec{\theta}=\gamma\mu_{0}(-\frac{\delta \vec{M}}{\chi_{M}}){+}\alpha\gamma\vec{\Gamma}'; \qquad
\label{dthetadtK=1}
\end{equation}
and using \eqref{dmdt0}, \eqref{Rm}, and \eqref{h*2},
\begin{equation}\partial_{t}\vec{m}-{D_m}
\nabla^{2}{\delta}\vec{m}=-\frac{\delta \vec{m}}{\tau_{m}}+\frac{\delta \vec{M}}{{\tau_{Mm}}}.
\label{j^mflow1}
\end{equation}
Eq.~\eqref{Gamma'2} should be used for $\vec{\Gamma}'$.

Consider the dc case, which is relevant for low-frequency studies using lock-in detectors.   Then \eqref{dMdtK=1} and \eqref{dthetadtK=1} can be used to eliminate $\vec{\Gamma}'$ and to obtain a linear relationship between $\delta\vec{M}$ and $\delta\vec{m}$, with a positive proportionality constant.  Then \eqref{j^mflow1} can be used to obtain an equation for $\vec{m}$, with a decay rate reduced from $\tau_{m}^{-1}$.  Despite the three equations, there is only a single doubly-degenerate diffusion mode, with a modified spin decay length $l_{SG}$.  (At finite frequency there also will be a non-diffusive mode that primarily involves $\delta\vec{M}$ and $\vec{\theta}$.)  


Note that the anisotropy $K$ can vary slowly with time.\cite{Pastora83}  In studying a dc spin current, we may take $K\rightarrow 0$.  Then, setting the time derivatives to zero in \eqref{dMdtK=1} and \eqref{dthetadtK=1} permits us to eliminate $\vec{\Gamma}'$ and to relate $\vec{M}$ and $\vec{m}$.  Placing this in \eqref{j^mflow1} then gives an effective wavevector squared $q^{2}$ for the diffusion mode, with $\vec{M}$ and $\vec{m}$ related.  $\vec{\Gamma}'$ then determines $\vec{\theta}$ because $\vec{\Gamma}'$ is related to $\vec{M}$ and $\vec{m}$, and includes the now-known wavevector $q^{2}$.

Using \eqref{deltaMm}, we now rewrite \eqref{dMdtK=1},  \eqref{dthetadtK=1}, and  \eqref{j^mflow1},
\begin{align}
&\partial_{t}\vec{M}=-\gamma\vec{\Gamma}'-\frac{\vec{M}+\vec{m}}{\tau_{M}}+\frac{\xi\vec{M}+\vec{m}}{{\tau_{mM}}},\\
&\partial_{t}\vec{\theta}=-\gamma\mu_{0} \frac{\vec{M}+\vec{m}}{\chi_{M}}+\alpha\gamma\vec{\Gamma}',\\
&\partial_{t}\vec{m}-{D_m}\nabla^{2}(\xi\vec{M}+\vec{m})=-\frac{\xi\vec{M}+\vec{m}}{\tau_{m}}+\frac{\vec{M}+\vec{m}}{{\tau_{Mm}}},
\end{align}
where
\begin{equation}
\vec{\Gamma}'=-K\vec{\theta}+\rho_{s}\nabla^{2}\vec{\theta},
\end{equation}
and $\vec{M}$ should be replaced by $\Delta \vec M=\vec{M}-\vec{M}_{0}-\chi_{M}\vec{H}$ if $\vec{M}_{0}-\chi_{M}\vec{H}\ne 0$.

We now assume that this physical system is subject to space- and time-variations of the form $e^{i (kx-\omega t)}$.  This can be done either by driving the system at frequency $\omega$, to which the system responds at some wavevector $k$ that must be determined, or by turning on a disturbance at a fixed $k$, to which the system responds in time at some $\omega$ (which may be complex) that must be determined. \\
\indent We will be particularly interested in fixed $\omega$, and determining $k$.  To that purpose it will be helpful to define
\begin{equation}
K'\equiv K+\rho_s k^2.
\label{K'}
\end{equation}
We also introduce
\begin{equation}
\frac{1}{\tau_{M}}\equiv\frac{1}{\tau_{ML}}+\frac{1}{\tau_{Mm}},  \quad \frac{1}{\tau_{m}}\equiv\frac{1}{\tau_{mL}}+\frac{1}{\tau_{mM}},
\label{taus}
\end{equation}
\begin{equation}
r\equiv \frac{1}{\tau_{mM}}-\frac{1}{\tau_{Mm}}, \quad\xi \equiv \frac{\tau_{mM}}{\tau_{Mm}}.
\label{}
\end{equation}


With these terms defined, the equations of motion can be rewritten in matrix form as
\begin{align}\label{eigen-equ}
&-i\omega
\left( \begin{array}{c}
M\\
m\\
\theta
\end{array} \right)=\Gamma\left( \begin{array}{c}
M\\
m\\
\theta
\end{array} \right),
\end{align}
where the matrix $\Gamma$ is
\begin{align}\label{Gamma}
& \Gamma=
\left( \begin{array}{ccc}
-\frac{1}{\tau_{ML}}  & -\frac{1}{\tau_{ML}}+r  & \gamma K'\\
-\xi (Dk^2 +\frac{1}{\tau_{mL}})  & - Dk^2 -\frac{1}{\tau_{mL}} -r  &0 \\
-\frac{\gamma\mu_0}{\chi_M} & -\frac{\gamma\mu_0}{\chi_M} & -\alpha\gamma K'
\end{array} \right).
\end{align}
We will solve the equations \eqref{eigen-equ}, which define an eigenvalue problem, using perturbation theory around the propagating spin-wave mode and about the diffusive mode. \\

\section{Weakly Coupled Modes}
In practice we expect the modes of this system to be neither non-interacting nor strongly coupled.  Rather we expect one mode to be primarily propagating and one mode to be primarily diffusive.  The goal of this section is to determine the extent to which the secondary degree of freedom of each mode is coupled into the primary degree of freedom, as in perturbation theory.

\subsection{Spin-Wave $M$-$\theta$ mode}
The zeroth order spin-wave mode is obtained by setting $m=0$.  Then
\begin{align}
&\partial_{t}\vec{M}=-\gamma\vec{\Gamma}'-\frac{\vec{M} }{\tau_{ML}},\\
&\partial_{t}\vec{\theta}=-\gamma\mu_{0} \frac{\vec{M}}{\chi_{M}}+\alpha\gamma\vec{\Gamma}'.
\end{align}
Assuming the variation $e^{ikx-\omega t}$, 
we have
\begin{align}
&-i\omega
\left( \begin{array}{c}
M\\
\theta
\end{array} \right)=\left( \begin{array}{cc}
-\frac{1}{\tau_{ML}}  &   \gamma K'\\
-\frac{\gamma\mu_0}{\chi_M} &  -\alpha\gamma K'
\end{array} \right)\left( \begin{array}{c}
M\\
\theta
\end{array} \right).
\end{align}\\
The equation determining the eigenvalues is
\begin{align}\label{}
&\det\left( \begin{array}{cc}
-\frac{1}{\tau_{ML}} +i\omega &   \gamma K'\\
-\frac{\gamma\mu_0}{\chi_M} &  -\alpha\gamma K'+i\omega
\end{array} \right)\nn\\
&=-\omega^2-i\omega\left( \frac{1}{\tau_{ML}}+\alpha\gamma K'\right)\nn\\
&+\left(\frac{\alpha}{\tau_{ML}} +\frac{\gamma\mu_0}{\chi_M}\right) \gamma K'=0,
\end{align}
whose solutions are
\begin{align}\label{}
&\omega=-\frac{i}{2}\left( \frac{1}{\tau_{ML}}+\alpha\gamma K'\right)\nn\\
&\pm \sqrt{-\frac{1}{4}\left( \frac{1}{\tau_{ML}}-\alpha\gamma K'\right)^2+ \frac{\gamma^2\mu_0 K'}{\chi_M}}.
\end{align}
We will consider the situation where $\omega $ is given. We then write $K'$ in terms of $\omega$:
\begin{align}\label{K'inomega}
K'=\frac{i\omega\left(-\frac{1}{\tau_{ML}} +i\omega\right)}{\gamma\left[\alpha\left(-\frac{1}{\tau_{ML}} +i\omega\right)-\frac{\gamma\mu_0}{\chi_M}\right]}.
\end{align}
Substituting this into the equation for $M$, we get
\begin{align}\label{}
\left(-\frac{1}{\tau_{ML}} +i\omega\right)M+\gamma \frac{i\omega\left(-\frac{1}{\tau_{ML}} +i\omega\right)}{\gamma\left[\alpha\left(-\frac{1}{\tau_{ML}} +i\omega\right)-\frac{\gamma\mu_0}{\chi_M}\right]}\theta=0,
\end{align}
so the unperturbed modes satisfy
\begin{align}\label{}
\theta=-\left[\alpha\left(-\frac{1}{\tau_{ML}} +i\omega\right)-\frac{\gamma\mu_0}{\chi_M}\right]M.
\end{align}

We next use the unperturbed solution of $M$ in the equation of $m$. We have:
\begin{align}\label{}
 &\left[-\xi (Dk^2 +\frac{1}{\tau_{mL}})  \right]M \nn\\
 &+ \left[-(Dk^2 +\frac{1}{\tau_{mL}}+r) +i\omega\right] m=0.
\end{align}
Using $k^2=(K'-K)/\rho_s$, we can find $m$ in terms $M$ and $\omega$:
\begin{align}\label{m-pert}
&m
=\frac{  \xi [\frac{D}{\rho_s}(K'-K) +\frac{1}{\tau_{mL}}] }{ -[\frac{D}{\rho_s}(K'-K) +\frac{1}{\tau_{mL}}+r] +i\omega }M,
\end{align}
with $K'$ given in \eqref{K'inomega} in terms of $\omega$. For this perturbation calculation to be valid, the values of the material parameters and $\omega$ should make the proportionality coefficient in \eqref{m-pert} to be much smaller than $1$.

\subsection{Diffusion mode from $m$}
We next consider the situation where the amplitude $m$ is much larger than those of $M$ and $\theta$. The unperturbed mode then correspond to the solution of the equation on $\partial_t m$ and $\partial_t \theta$, with $M$ set to zero:
\begin{align}
&\partial_{t}\vec{m}-{D_m}\nabla^{2}\vec{m}=-\frac{ \vec{m}}{\tau_{m}}+\frac{ \vec{m}}{{\tau_{Mm}}}.
\end{align}
Substituting in the wave form of the solutions, we have
\begin{align}
(- Dk^2 -\frac{1}{\tau_{mL}} -r +i\omega )m=0.
\end{align}
We will consider the situation where $\omega $ is given. We then write $k^2$ in terms of $\omega$:
\begin{align}\label{k2inomega}
k^2=\frac{1}{D}(-\frac{1}{\tau_{mL}} -r +i\omega ).
\end{align}
We thus have
\begin{align}\label{K'inomegadiffusion}
K'=\rho_0+\frac{\rho_s  }{D}(-\frac{1}{\tau_{mL}} -r +i\omega).
\end{align}

We next use the unperturbed solution of $m$ in the equation of $M$ and $\theta$. We have:
\begin{align}\label{}
&(-\frac{1}{\tau_{ML}}+i\omega) M   +  \gamma K'\theta=-( -\frac{1}{\tau_{ML}}+r)m,\\
&(-\frac{\gamma\mu_0}{\chi_M} )M +(-\alpha\gamma K'+i\omega)\theta=\frac{\gamma\mu_0}{\chi_M} m ,
\end{align}
whose solution is:
\begin{align}\label{}
&M
=-\frac{( -\frac{1}{\tau_{ML}}+r)(-\alpha\gamma K'+i\omega)+\frac{\gamma^2\mu_0}{\chi_M}K'}{(-\frac{1}{\tau_{ML}}+i\omega)(-\alpha\gamma K'+i\omega)+  \frac{\gamma^2\mu_0}{\chi_M} K' }m,\\
&\theta=\frac{(-r+i\omega)\frac{\gamma\mu_0}{\chi_M} }{(-\frac{1}{\tau_{ML}}+i\omega)(-\alpha\gamma K'+i\omega)+\frac{\gamma^2\mu_0}{\chi_M} K' }m,
\end{align}
with $K'$ given in \eqref{K'inomegadiffusion} in terms of $\omega$. For this perturbation calculation to be valid, the values of the material parameters and $\omega$ should make the proportionality coefficients in these two equations much smaller than one.

The total magnetization $M+m$ for this perturbation solution, which is what would be measured, is given by:
\begin{align}\label{}
&M+m=(1+\frac{M}{m})m\nn\\
&=\left[1-\frac{( -\frac{1}{\tau_{ML}}+r)(-\alpha\gamma K'+i\omega)+\frac{\gamma^2\mu_0}{\chi_M}K'}{(-\frac{1}{\tau_{ML}}+i\omega)(-\alpha\gamma K'+i\omega)+  \frac{\gamma^2\mu_0}{\chi_M} K' }\right]m,
\end{align}
with $K'$ given in \eqref{K'inomegadiffusion}.

\section{Summary and Discussion}

We have developed the theory of spin diffusion in spin glasses, finding it necessary to invoke the spin accumulation $\vec{m}$, due to a non-equilibrium distribution of excitations.  Although prompted by experiments, at the moment there is a need for further experiments in order to compare with the general predictions of the theory, and to determine parameters appearing in the theory.

The present theory was developed with insulators in mind.  A metallic spin glass, such as Cu doped with Mn, is more complex.\cite{Mydosh15}   There the randomly located Mn has spin-glass order, with exchange between two localized Mn spins at irregular positions having irregular sign.  This exchange is believed to be due to the RKKY interaction, which is mediated by the Cu host conduction electrons that communicate between the two Mn (See Appendix A on the RKKY interaction.) In addition, however, each Mn spin-polarizes the bath of Cu electrons in its vicinity, thus making the host Cu a (likely weak) itinerant spin-glass that responds to the Mn spin-order.

For a spin current to propagate through such a system, there must be a spin current in both the Cu and the Mn.  Spin currents in Cu are carried by delocalized conduction electrons by means of a spatially-varying non-equilibrium spin-distribution function.  Spin currents in Mn are carried {by} localized Mn by means of a spatially-varying non-equilibrium magnon-distribution function.  This is a rather complex situation, likely requiring at the microscopic level a theory for the spin current that is spatially averaged over both the Cu and the Mn.  That is well beyond the scope of the present work.

However, the present macroscopic theory serves the purpose of describing macroscopic spin currents even in CuMn.  It does not consider the possibility of a charge current, as can occur in the conductor CuMn, as opposed to the insulator YIG.


\acknowledgements{}
C. S. is supported by the Fundamental Research Funds for the Central Universities from China.

\vskip0.5cm
\appendix
\section{On the RKKY Interaction}
The RKKY interaction was initially developed for collinear nuclear spins interacting indirectly with conduction electrons. In order to obtain a mechanism for the line {\it broadening} observed for {\it nuclei} in nuclear magnetic resonance, Ruderman and Kittel considered indirect exchange mediated by the hyperfine interaction between conduction electrons and nuclei.\cite{RudermanKittel54}  It took the form $A_{ij}\vec{I}_{i}\cdot\vec{I}_{j}$ between nuclear spins $\vec{I}_{n}$.  With $k_{F}$ the host Fermi wave vector and $R_{ij}$ the separation between two nuclei, they found an oscillatory and inverse power law dependence of the exchange constant $A_{ij}$ on $2k_{F}R_{ij}$.  They noted that although exchange {\it narrowing} will occur for pure isotopic samples, for naturally occurring isotopic mixtures such an interaction (being inhomogenous) can lead to the observed line {\it broadening}.


Kasuya, noting Zener's phenomenology on the interaction of core electrons (d states) via the s-d interaction with s-state conduction electrons,\cite{Zener51} then modified the Ruderman and Kittel approach to treat d-d electron interactions, now mediated by the s-d interaction.\cite{Kasuya56}  Yosida applied these ideas to CuMn alloys,\cite{Yosida57} noting that ``the experimental results on the electronic $g$-value of the Mn ions and the Knight shift of the Cu-nuclei can be qualitatively accounted for.''   Later experiments on CuMn alloys at low temperatures indicated a complex magnetic structure.  Van Vleck provided a clarifying discussion of theoretical details,\cite{VanVleck62} and refered to an earlier paper by Fr\"ohlich and Nabarro\cite{FrohlichNabarro40} that contains an exchange interaction between nuclear spins with no dependence on their relative position.  Van Vleck also cites additional papers where Zener\cite{Zener51-52} expands on his earlier work.\cite{Zener51}

It was later found that when the Mn spins are permitted to be noncollinear, the spacially oscillating interaction between randomly placed Mn impurities minimizes the Mn energy for a seemingly random non-collinear spin structure called a spin-glass.\cite{EdwardsAnderson75}

{}


\begin{thebibliography}{}

\bibitem{Wolf05} S. A. Wolf, D. D. Awschalom, R. A. Buhrman, J. M. Daughton, S. von Molnar, M. L. Roukes, A. Y. Chtchelkanova, and D. M. Treger, Science 294, 1488 (2005). ``Spintronics: a spin-based electronics vision for the future''.


\bibitem{ZFD-RMP} I. \v{Z}uti\'{c}, J. Fabian, and S. Das Sarma, Rev. Mod. Phys. 76, 323 (2004). ``Spintronics: Fundamentals and applications''.


\bibitem{Iguchi10} R. Iguchi, K. Ando, E. Saitoh, T .Sato, J. Phys: Conf. Series 266, 012089 (2011).  ``Spin current study of spin glass AgMn using spin pumping effect.''


\bibitem{NatPhysWuZink17} D. Wesenberg, T. Liu, D. Balzar, M. Wu and B. L. Zink, Nat. Phys. 13, 987 (2017). ``Long-distance spin transport in a disordered magnetic insulator''.

\bibitem{Mydosh15} J. A. Mydosh, Rep. Prog. Phys. 78, 052501 (2015). ``Spin glasses: redux: an updated experimental/materials survey''.


\bibitem{WangHammel14} H. Wang, C. Du, P. C. Hammel, F. Yang, Phys. Rev. Lett. 113, 097202 (2014). ``Antiferromagnonic Spin Transport from Y$_3$Fe$_5$O$_{12}$ into NiO''.

\bibitem{Rezende16} S. M. Rezende, L. Rodr\'{i}guez-Su\'{a}rez, and A. Azevedo, Phys. Rev.  B 93, 054412 (2016). ``Diffusive magnonic spin transport in antiferromagnetic insulators.''

\bibitem{WalkerWalstedt77} L. R. Walker and R. E. Walstedt, Phys. Rev. Lett. 38, 514 (1977).  ``Computer Model of Metallic Spin Glasses''.

\bibitem{WalkerWalstedt80} L. R. Walker and R. E. Walstedt, Phys. Rev. B 22, 3816 (1980).  ``Computer Model of Metallic Spin Glasses''.

\bibitem{HalperinSaslow77} B. I. Halperin and W. M. Saslow, Phys. Rev. B 16, 2154 (1977).  ``Hydrodynamic theory of spin waves in spin glasses and other systems with noncollinear spin orientations''.

\bibitem{Andreev78} A.~F. Andreev, Sov. Phys. JETP 47, 411 (1978). ``Magnetic properties of disordered media''.

\bibitem{ValetFert93} T. Valet and A. Fert, Phys. Rev. B 48, 7099 (1993).  ``Theory of the perpendicular magnetoresistance in magnetic multilayers''.

\bibitem{ZhangLevyFert02} S. Zhang, P. M. Levy, and A. Fert, Phys. Rev. Lett. 88, 236601 (2002).  ``Mechanisms of Spin-Polarized Current-Driven Magnetization Switching''.

\bibitem{DPJETPLett71} M. I. Dyakonov and V. I. Perel, Sov. Phys. JETP Lett. 13, 467 (1971). ``Possibility of Orienting Spins with Current''.

\bibitem{DPPhysLett71} M. I. Dyakonov and V. I. Perel, Phys. Lett. A 35, 459 (1971). ``Current-induced Spin Orientation of Electrons in Semiconductors''.

\bibitem{SasSunXu22}  W. M. Saslow, C. Sun, and S. Xu, Phys. Rev. B 105, 174441 (2022).  ``Spin accumulation and longitudinal spin diffusion of magnets''.

\bibitem{JohnsonSilsbee87a} M. Johnson and R. H. Silsbee, Phys. Rev. B 35, 4959 (1987).  ``Thermodynamic analysis of interfacial transport and of the thermomagnetoelectric system''.

\bibitem{Aranov76} A. G. Aronov,  Sov. Phys. JETP Lett. 24, 32 (1976).  ``Spin Injection in metals and polarization of nuclei''.

\bibitem{Wyder87} P. C. van Son, H. van Kempen, and P. Wyder, Phys. Rev. Lett. 58, 2271 (1987). ``Boundary Resistance of the Ferromagnetic-Nonferromagnetic Metal Interface''.

\bibitem{LevyMPFert82} P. M. Levy, C. Morgan-Pond, and A. Fert, J. Appl. Phys. 53, 2168 (1982). ``Origin of anisotropy in transition metal spin glass alloys''.

\bibitem{HSH82} C. L. Henley, H. Sompolinsky, and B. I. Halperin, Phys. Rev. B 25, 5849 (1982). ``Spin-resonance frequencies in spin-glasses with random anisotropies''.

\bibitem{WMS83} W. M. Saslow, Phys. Rev. B 27, 6873 (1983).  ``Anisotropy and anisotropy-triad dynamics in spin-glasses''.


\bibitem{Rahman85} A. Rahman, summer 1985.  He and the author were in a centralized computer room at Argonne National Labs, respectively performing simulations on real glasses and spin glasses.  For him the context of an ``ether peg'' was to seed (symmetry-break) a liquid-to-solid transition around a known point in space.  The Computational Physics Prize of the American Physical Society is aptly named after Rahman.

\bibitem{Pastora83} J. B. Pastora, D. Love, and T. W. Adair, J. de Physique Lett., 44, 859-863 (1983).  ``Anisotropy dynamics of CuMn spin glass through torque measurements''.

\bibitem{WMSDP} W. M. Saslow, Phys. Rev. B 91, 014401 (2015).  ``Spin Hall effect and irreversible thermodynamics: Center-to-edge transverse current-induced voltage.''

\bibitem{WMS82} W. M. Saslow, Phys. Rev. Lett. 48, 505 (1982). ``Anisotropy and anisotropy-triad dynamics in spin-glasses''.

\bibitem{RudermanKittel54} M. A. Ruderman and C. Kittel, Phys. Rev. 96, 99 (1954). ``Indirect Exchange Coupling of Nuclear Magnetic Moments by Conduction Electrons''.

\bibitem{Zener51} C. Zener, Phys. Rev. 81, 440 (1951). ``Interaction Between the d Shells in the Transition Metals''.

\bibitem{Kasuya56} T. Kasuya, Prog. Th. Phys. 16, 45 (1956). ``A Theory of Metallic Ferro- and Antiferromagnetism on Zener's Model''.

\bibitem{Yosida57} K. Yosida, Phys. Rev. 106, 893 (1957). ``Magnetic Properties of Cu-Mn Alloy''.

\bibitem{VanVleck62} J. H. Van Vleck, Rev. Mod. Phys. 34, 681.(1962). ``Note on the Interactions between the Spins of Magnetic Ions or Nuclei in Metals''.

\bibitem{FrohlichNabarro40} H. Frohlich and F. R. N. Nabarro, Proc. Roy. Soc. (London) A175, 382 (1940).  ``Orientation of nuclear spins in metals''.

\bibitem{Zener51-52}
C. Zener, Phys. Rev. 82, 403 (1951). ``Interaction between the d-Shells in the Transition Metals. II. Ferromagnetic Compounds of Manganese with Perovskite Structure''.
C. Zener, Phys. Rev. 83, 299 (1951). ``Interaction between the d-Shells in the Transition Metals.  III. Calculation of the Weiss Factors in Fe, Co, and Ni''.
C. Zener, Phys. Rev. 85, 324 (1952). ``Interaction between the d-Shells in the Transition Metals.  IV. The Intrinsic Antiferromagnetic Character of Iron''.

\bibitem{EdwardsAnderson75} S.~F. Edwards and P.~W. Anderson, J. Phys. F: Met. Phys. 5, 965 (1975). ``Theory of spin glasses''.

\end{thebibliography}
\end{document}